\documentclass[11pt,a4paper]{article}

\usepackage[T1]{fontenc}
\usepackage[utf8]{inputenc}
\usepackage{lmodern}
\usepackage{amsmath,amssymb,bm}
\usepackage{graphicx}
\usepackage{microtype}
\usepackage{cite}
\usepackage[hidelinks]{hyperref}
\usepackage[margin=2.15cm]{geometry}

\newcommand{\Id}{\mathbb{I}}
\newcommand{\sz}{\sigma_z}
\newcommand{\ve}{\nu_e}
\newcommand{\vmu}{\nu_\mu}
\newcommand{\dm}{\delta m}
\newcommand{\dmsq}{\Delta m^2}

\title{\textbf{Concurrence Dynamics in Self-Gravitating Mixed Neutrinos}}

\author{
Raoul Serao$^{1}$, Gianpaolo Torre$^{2}$, Antonio Capolupo$^{3}$\\[0.45em]
\small $^{1}$Department of Physics, Faculty of Science, University of Zagreb, Zagreb, Croatia\\
\small $^{2}$Jožef Stefan Institute, Ljubljana, Slovenia\\
\small $^{3}$Dipartimento di Fisica ``E.R. Caianiello'', Università di Salerno,\\
\small and INFN -- Gruppo Collegato di Salerno, Fisciano, Italy
}

\date{}

\begin{document}

\maketitle

\begin{abstract}
We study the entanglement generated by Newtonian self-gravity in a two-particle
system of mixed neutrinos and quantify it by means of the concurrence.
Starting from the gravitational coupling between the mass operators, we show
that the two-particle Hamiltonian splits into local single-particle terms and
an effective Ising interaction in the mass basis, the latter being solely
responsible for the generation of correlations.
For initially separable flavor states the concurrence is obtained in closed
analytic form: it oscillates between zero and a maximum fixed by the mixing
angle alone, while the gravitational coupling sets the recurrence time.
We further establish a one-to-one correspondence between the concurrence and
the reduced-state purity previously derived for this model, thus providing an
independent quantum-information description of the same gravitationally
induced correlations.
Finally, we show that the concurrence does not depend on the Majorana phase,
so that Dirac and Majorana parametrizations yield identical entanglement
dynamics.
These results isolate the genuinely entangling sector of the gravitational
evolution and provide a compact analytic characterization of gravity-induced
quantum correlations in mixed neutrino systems.
\end{abstract}

\section{Introduction}
\label{sec:introduction}
%% ---- STATO DELLE NOTE EDITORIALI ----
%% FATTO: "it has been shown in Ref." al posto dei nomi degli autori (fine Sec. I)
%% FATTO: eliminate le citazioni 20, 21, 26
%%        (Blasone2007vw, Blasone2015lya, Blasone2021cau)
%% NON FATTO (di proposito): eq. (10), il quadrato NON va eliminato: con
%%        ad-bc = (i/2) sin^2(2theta) sin(2 Omega t) si ha C = sin^2(2theta)|sin(2 Omega t)|,
%%        unico risultato compatibile con la purity via C^2 = 2(1-P).
%% DA FARE: inserire altri lavori su curved background prima della ref. 14
%%        (CapolupoSerao2023) -> servono le chiavi di references.bib
%% DA CONFERMARE: portare nel testo le eq. 3, 5, 12, 13, 14, 16 (16 come diag(1,e^{i\phi}))
%%        e mettere su un solo rigo le eq. 6 e 9
Neutrino mixing is one of the clearest manifestations of quantum coherence in
particle physics. The oscillation framework is by now firmly established
experimentally, and describes flavor conversion as the result of the coherent
superposition of states with different masses
\cite{BilenkyPontecorvo1978,BilenkyPetcov1987,Fukuda1998,Ahmad2001,
PDG2024,GiuntiKim2007}.
At the field-theoretical level, however, particle mixing displays a richer
structure, which includes a nontrivial flavor vacuum and corrections that are
absent in the usual first-quantized description \cite{BlasoneVitiello1995,BlasoneCapolupoVitiello2002,Fujii1999,
Fujii2001,Hannabuss2000,BlasoneCapolupoRomeiVitiello2001,
CapolupoCapozzielloVitiello2007}.
The same framework has been extended to curved and torsional backgrounds,
where the geometry can affect the mixing transformation, the oscillation
frequencies, and the structure of the flavor condensate
\cite{Curv1, Curv2, Curv3,Curv4,CapolupoSerao2023,Capolupo2023igw,Antonio2026asy,Serao2026ntn,
Capolupo2019gmn,CapolupoSerao2025DarkUniverse}.

The superposition structure of flavor states also makes it possible to define
and study quantum coherence, entanglement, Bell-type correlations,
complementarity relations, and geometric phases in a high-energy setting
\cite{Alok2014gya,Banerjee2015mha,Dixit2019lsg,Ettefaghi2020otb,
Bittencourt2022tcl,Capolupo2018hrp,Buoninfante2020iyr,
CapolupoGiampaolo2020,Serao2025xqw,Capolupo2026nuy}.
While these features already follow from the coherent structure of flavor
mixing alone, interactions between distinct neutrinos open a qualitatively
different regime, in which entanglement is dynamically generated between
initially separable particles.

In this regime the evolution becomes genuinely many-body, and a mean-field
factorization neglects dynamically generated correlations. Exact solutions,
tensor-network calculations, and quantum simulations have therefore been
employed to quantify such effects
\cite{Pantaleone1992,SiglRaffelt1993,Duan2010bg,Pehlivan2011hp,
Patwardhan2019zta,Cervia2019res,Roggero2021asb,Rrapaj2020,
Hall2021rbv,Cervia2022pro,Siwach2022xhx,Capolupo2020enx}.
More generally, quantum-information observables are increasingly used to recast
questions concerning weakly coupled sectors and fundamental symmetries in terms
of interferometric or correlation-based probes
\cite{Capolupo2025zei,Capolupo2024xlp,Serao2026hcp,Serao2026zeu}.

A setting in which this mechanism can be analyzed analytically is provided by
self-gravitating mixed neutral particles.
It has been shown that, within a Newtonian model at fixed interparticle
separation, the gravitational interaction between particles with mixed masses
can generate entanglement out of initially separable states and modify the
corresponding flavor-transition probabilities
\cite{Simonov2019}.
The resulting dynamics can be mapped onto an effective Ising-type interaction,
which provides a direct connection with quantum spin models and allows the same
structure to be explored in analogue quantum simulators, including trapped-ion
platforms \cite{Capolupo2021Trapped}.
The use of entanglement as a probe of weak interactions also relates this model
to proposals for witnessing gravity-mediated correlations and to analyses of
gravitational decoherence
\cite{Bose2017nin,Marletto2017kzi,Kafri2014zsa,Carney2018ofe,
Bassi2017szd,Christodoulou2018cmk,Pikovski2011zk,Albrecht2014sxa,
Howl2018qdl,GiampaoloMacri2019,CapolupoAxion2020}.
In the original formulation, the generated entanglement was quantified through
the purity of the reduced single-particle state and the corresponding
R\'enyi entropy.

In this work we revisit the two-particle sector of this model for two-flavor
neutrino mixing, and characterize the gravitationally generated entanglement
directly in terms of the concurrence.
Unlike the reduced-state purity, the concurrence is an entanglement monotone
normalized to the interval $[0,1]$ which, for two-qubit systems, admits a
closed-form expression
\cite{HillWootters1997,Wootters1998,Horodecki2009,Rungta2001,Amico2008}.
Separability, partial entanglement, and maximal entanglement can therefore be
identified directly from its value. The aim is thus not to introduce a
different gravitational dynamics, but to extract a more direct
quantum-information description of the correlations already generated by the
underlying interaction.

We first derive the two-particle gravitational Hamiltonian from the Newtonian
interaction between the mass operators of the two mixed particles.
Once the state-independent contribution is removed, the Hamiltonian separates
into local single-particle terms and an interaction proportional to
$\sigma_z\otimes\sigma_z$.
This decomposition makes the origin of the entanglement explicit: the local
terms only modify the phases of the individual particles and leave the
concurrence unchanged, whereas the Ising-like gravitational term acts as a
conditional-phase interaction and is entirely responsible for the generation of
bipartite entanglement.

For initially separable flavor states we then obtain the concurrence in closed
analytic form. The result describes a periodic creation and disappearance of
entanglement, and separates two physically distinct ingredients of the
dynamics: the mixing angle determines the largest concurrence that can be
attained, while the gravitational coupling fixes the time scale over which
entanglement is generated and subsequently recurs. The conditions for maximal
entanglement, the times at which the maxima occur, and the recurrence period
follow explicitly.
We further connect the concurrence with the reduced-state purity of
Refs.~\cite{Simonov2019}, and show that the two descriptions encode
the same bipartite correlations for the closed two-particle system.
Finally, we analyze the role of the Majorana phase and show that, for the
Hamiltonian considered here, Dirac and Majorana parametrizations lead to the
same concurrence dynamics, since they are related by local phase
transformations.

The paper is organized as follows.
In Sec.~\ref{sec:model} we introduce the two-flavor neutrino states and derive
the effective two-particle Hamiltonian from the Newtonian interaction between
the mass operators.
In Sec.~\ref{sec:concurrence} we isolate the entangling part of the evolution
and derive the time-dependent concurrence for initially separable flavor
states.
In Sec.~\ref{sec:purity} we establish the relation between concurrence,
reduced-state purity, and R\'enyi entropy, discuss the maximal attainable
entanglement together with the associated characteristic and recurrence times,
and compare the Dirac and Majorana parametrizations, showing the absence of any
Majorana-phase dependence in the concurrence.
Section~\ref{sec:conclusions} summarizes the results and discusses possible
extensions beyond the two-particle sector.

\section{Two-flavor gravitational model}
\label{sec:model}
We work in natural units, $\hbar=c=1$, and adopt the mass basis
$\{|m_1\rangle,|m_2\rangle\}$ as the computational basis.
The two-flavor states read
\begin{align}
|\ve\rangle
&=
\cos\theta\,|m_1\rangle
+
e^{i\phi}\sin\theta\,|m_2\rangle,
\\
|\vmu\rangle
&=
-e^{-i\phi}\sin\theta\,|m_1\rangle
+
\cos\theta\,|m_2\rangle ,
\label{eq:flavorstates}
\end{align}
where $\theta$ is the mixing angle and $\phi$ is the Majorana phase, the Dirac
case being recovered for $\phi=0$.
Standard flavor-oscillation probabilities are insensitive to the Majorana phase
\cite{Giunti2010}.

For ultrarelativistic neutrinos of common energy $E\gg m_i$, the single-particle
energies can be expanded as $E_i\simeq E+m_i^2/2E$, so that the traceless
(state-dependent) part of the free Hamiltonian is
$H^{(1)}=\omega_0\sz,
\quad
\omega_0=
\frac{m_1^2-m_2^2}{4E}.$
For two particles the free contribution is accordingly
$H_0^{(2)}
=
\omega_0
\left(
\sz\otimes\Id+\Id\otimes\sz
\right).$

The gravitational contribution is derived below rather than introduced
phenomenologically. Assuming that gravity couples to the mass operator, which
in the mass basis reads
\begin{equation}
\widehat M
=
\frac{m_1+m_2}{2}\Id
+
\frac{m_1-m_2}{2}\sz,
\label{eq:massop}
\end{equation}
and working in the Newtonian regime at fixed interparticle separation $d$, the
interaction operator between the two particles is
$V_G
=
-\frac{G}{d}
\,\widehat M_A\widehat M_B,
$
where the subscripts $A$ and $B$ label the first and the second particle,
respectively.
Treating $d$ as a fixed external parameter implies that wave-packet separation,
the motion of the particles, relativistic corrections to the gravitational
interaction, and backreaction are not included in the model.

Introducing the mass splittings
$\dm\equiv m_1-m_2,
\quad
\dmsq\equiv m_1^2-m_2^2,$
we have
\begin{equation}
V_G
={}
-\frac{G}{d}
\left[
\frac{(m_1+m_2)^2}{4}\Id\otimes\Id
+
\frac{\dmsq}{4}
\left(
\sz\otimes\Id+\Id\otimes\sz
\right)
\right.
\left.
+
\frac{(\dm)^2}{4}
\sz\otimes\sz
\right].
\label{eq:VGexpanded}
\end{equation}
The first term is proportional to the identity and contributes only a global
phase. Discarding this state-independent contribution and defining
$g=-\frac{G}{4d},$
the dynamically relevant Hamiltonian reads
\begin{equation}
H^{(2)}
=
\omega
\left(
\sz\otimes\Id+\Id\otimes\sz
\right)
+
\Omega\,\sz\otimes\sz,
\label{eq:H2}
\end{equation}
with
$\omega
=
\omega_0+g\,\dmsq,
\Omega
=
g\,(\dm)^2.$
Equation~\eqref{eq:H2} coincides with the two-particle gravitational
Hamiltonian used in Ref.~\cite{Simonov2019}.
The two terms are controlled by different mass combinations: the local
gravitational shift is proportional to $m_1^2-m_2^2$, whereas the entangling
coupling is proportional to $(m_1-m_2)^2$ and is therefore quadratically
suppressed for nearly degenerate masses.

\section{Interaction-induced concurrence}
\label{sec:concurrence}
The two terms in Eq.~\eqref{eq:H2} commute, so that the evolution operator
factorizes,
$U(t)
=
U_{\mathrm{loc}}(t)\,
U_{\mathrm{int}}(t),$
where
$U_{\mathrm{loc}}(t)
=
e^{-i\omega t\sz}
\otimes
e^{-i\omega t\sz},
U_{\mathrm{int}}(t)
=
e^{-i\Omega t\sz\otimes\sz}.$
The first factor is a product of local unitaries and therefore leaves the
two-particle concurrence unchanged.
The entangling dynamics is thus generated entirely by the gravitational
$\sz\otimes\sz$ term.

Consider the initially separable state
$|\Psi(0)\rangle
=
|\ve\rangle\otimes|\vmu\rangle.$
For definiteness we first adopt the Dirac convention $\phi=0$ and set
$c=\cos\theta$, $s=\sin\theta$; the role of $\phi$ is analyzed in
Sec.~\ref{sec:majorana}.
In the computational basis,
\begin{equation}
|\Psi(0)\rangle
=
-cs|00\rangle
+c^2|01\rangle
-s^2|10\rangle
+cs|11\rangle.
\end{equation}
Up to local phases, the interaction evolves this state into
\begin{equation}
|\Psi(t)\rangle
={}
-cs\,e^{-i\Omega t}|00\rangle
+c^2 e^{i\Omega t}|01\rangle
-
s^2 e^{i\Omega t}|10\rangle
+cs\,e^{-i\Omega t}|11\rangle.
\label{eq:psit}
\end{equation}
For a pure two-qubit state
$|\psi\rangle
=
\alpha|00\rangle+\beta|01\rangle+\gamma|10\rangle+\delta|11\rangle,$
the concurrence is
$C=2|\alpha\delta-\beta\gamma|$
\cite{Wootters1998}.
Equation~\eqref{eq:psit} yields
$\alpha\delta-\beta\gamma
=
\frac{i}{2}
\sin^2(2\theta)
\sin(2\Omega t),$
and therefore
\begin{equation}
C(t)
=
\sin^2(2\theta)
\left|
\sin(2\Omega t)
\right|
=
\sin^2(2\theta)
\left|
\sin\!\left[
\frac{G(m_1-m_2)^2}{2d}\,t
\right]
\right|
.
\label{eq:Cgeneric}
\end{equation}
Only the modulus enters, so that the sign of $g$ is immaterial for the
concurrence.

\begin{figure}[t]
\centering
\includegraphics[width=0.86\linewidth]{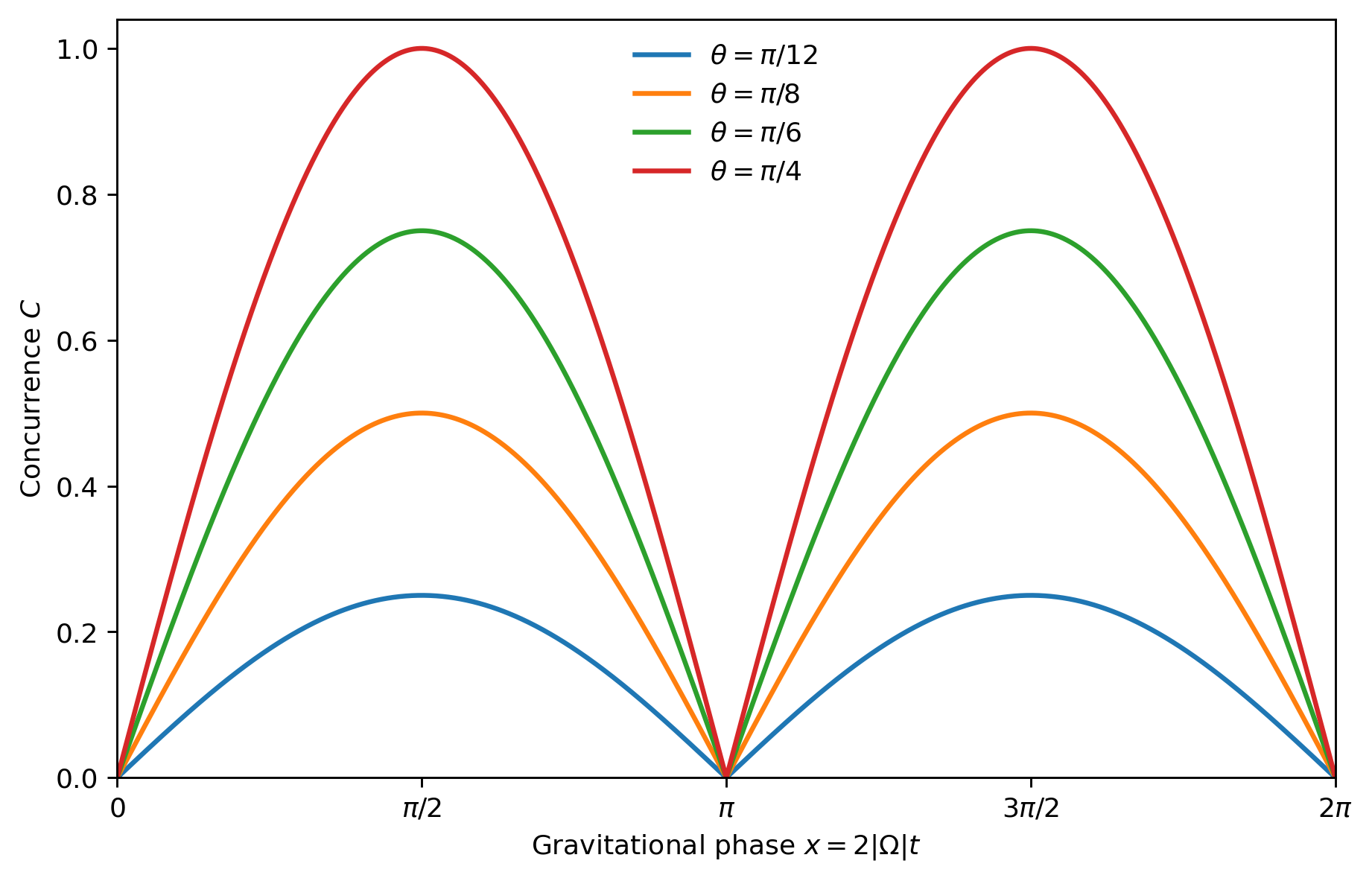}
\caption{
Concurrence as a function of the dimensionless gravitational phase
$x=2|\Omega|t$ for representative values of the mixing angle.
The maximal value is $C_{\max}=\sin^2(2\theta)$.
}
\label{fig:concurrence}
\end{figure}

Figure~\ref{fig:concurrence} displays Eq.~\eqref{eq:Cgeneric} as a function of
the dimensionless gravitational phase $x=2|\Omega|t$ for representative values
of the mixing angle. The mixing angle controls the amplitude of the
oscillation, while gravity fixes its period.

The result does not depend on the particular separable flavor product chosen as
initial state: the choices $|\ve\ve\rangle$, $|\ve\vmu\rangle$,
$|\vmu\ve\rangle$, and $|\vmu\vmu\rangle$ all yield the same concurrence.

\section{Purity, time scales, and the Majorana phase}
\label{sec:purity}

\subsection{Relation to the reduced-state purity}
\label{sec:purityrel}
In the original gravitational analysis, entanglement was quantified through the
purity of the reduced state of either particle, which for the two-particle
system reads
\begin{equation}
\mathcal P(t)
=
1-
\frac{1}{2}
\sin^4(2\theta)
\sin^2(2\Omega t)
\label{eq:purity}
\end{equation}
\cite{Simonov2019}.
For a pure state of two qubits, concurrence and reduced purity are related by
\begin{equation}
C^2=2(1-\mathcal P),
\label{eq:Cpurity}
\end{equation}
i.e. the tangle equals twice the linear entropy of the reduced state.
Substituting Eq.~\eqref{eq:purity} into Eq.~\eqref{eq:Cpurity} reproduces
Eq.~\eqref{eq:Cgeneric}, so that the present calculation is independently
consistent with the previous purity result.

The R\'enyi-2 entanglement entropy used in Ref.~\cite{Simonov2019},
$S_2=-\ln\mathcal P,$
can likewise be expressed in terms of the concurrence as
$S_2
=
-\ln\left(1-\frac{C^2}{2}\right).$
The three quantities therefore carry equivalent information for this closed
pure two-particle problem, the concurrence having the advantage of a direct
normalization between $0$ and $1$.

\subsection{Characteristic and recurrence times}
\label{sec:times}
Equation~\eqref{eq:Cgeneric} makes the relevant scales transparent.
The maximum attainable concurrence is
$C_{\max}
=
\sin^2(2\theta),$
so that maximal bipartite entanglement, $C=1$, is reached only for maximal
mixing,
$\theta=\frac{\pi}{4}.$
Irrespective of the value of $\theta$, the maxima of the concurrence occur at
\begin{equation}
t_{\max}^{(k)}
=
\frac{(2k+1)\pi}{4|\Omega|}
=
\frac{(2k+1)\pi d}
{G(m_1-m_2)^2},
\qquad
k=0,1,2,\ldots
\label{eq:tmax}
\end{equation}
while the concurrence vanishes at
\begin{equation}
t_{0}^{(k)}
=
\frac{k\pi}{2|\Omega|}
=
\frac{2k\pi d}
{G(m_1-m_2)^2}.
\end{equation}
The recurrence period of the concurrence is therefore
\begin{equation}
T_C
=
\frac{\pi}{2|\Omega|}
=
\frac{2\pi d}
{G(m_1-m_2)^2}.
\label{eq:period}
\end{equation}
These expressions show explicitly why the effect is extremely slow for
realistic neutrino masses and macroscopic separations, while retaining a clean
conceptual interpretation.

\subsection{Dirac and Majorana parametrizations}
\label{sec:majorana}
For the convention adopted in Eq.~\eqref{eq:flavorstates}, we define the local
phase operator
\begin{equation}
D_\phi
=
\begin{pmatrix}
1&0\\
0&e^{i\phi}
\end{pmatrix}.
\end{equation}
The electron-flavor Majorana state is obtained from the corresponding Dirac
state through $D_\phi$, and the muon-flavor state differs by the same local
phase transformation up to an irrelevant global phase.
Moreover,
$[
H^{(2)},D_\phi\otimes D_\phi
]=0,$
since both operators are diagonal in the mass basis.
The Dirac and Majorana two-particle states therefore remain related by local
unitary transformations at all times, and, the concurrence being invariant
under local unitaries,
\begin{equation}
C_M(t)=C_D(t).
\end{equation}

The absence of a Majorana-phase dependence is thus structural rather than the
result of an accidental cancellation.
For this gravitational Ising interaction the concurrence quantifies the
generated particle--particle entanglement, but cannot discriminate between the
Dirac and Majorana nature of neutrinos.
This is consistent with the phase independence of the flavor-transition
probabilities obtained within the same model \cite{Simonov2019}.

\section{Conclusions}
\label{sec:conclusions}

We have provided a direct quantum-information characterization of the
entanglement generated by Newtonian self-gravity in a two-particle system of
mixed neutrinos. Building on the self-gravitating mixing framework of
Refs.~\cite{Simonov2019}, we have shown that the gravitational
dynamics admits a transparent decomposition into local single-particle
contributions and a genuinely entangling interaction. This separation allows
one to identify unambiguously which part of the gravitational evolution is
responsible for the creation of correlations between initially separable
particles.

The resulting concurrence dynamics yields a simple physical picture of the
gravitationally induced entanglement. The local gravitational contribution
affects only single-particle phases and cannot modify the amount of bipartite
entanglement, whereas the interaction term acts as a conditional-phase coupling
in the mass basis and periodically drives the system between separable and
entangled configurations. Two distinct ingredients control this dynamics: the
flavor-mixing angle fixes the maximum amount of entanglement that can be
generated, while the gravitational coupling determines the time scale over
which entanglement is created, lost, and subsequently revived. The concurrence
therefore separates the kinematical role of mixing from the dynamical role of
the interaction.

We have also established an exact correspondence between the concurrence and
the reduced-state purity previously obtained for the same model. The
concurrence does not introduce a different notion of the underlying
correlations, but provides a normalized entanglement monotone from which the
conditions for separability, partial entanglement, maximal entanglement, and
recurrence can be read directly. This connects the original
reduced-density-matrix description to a standard two-qubit quantum-information
framework.

Furthermore, the concurrence has been shown to be insensitive to the Majorana
phase for the gravitational Hamiltonian considered here. This is not a
consequence of a special choice of parameters or of an accidental cancellation:
Dirac and Majorana parametrizations are related by local phase transformations
that commute with the evolution, and the concurrence is invariant under local
unitary operations. The equality of the two entanglement dynamics is thus a
structural property of the model, and clarifies which quantum-information
observables can, and cannot, carry information on the nature of the neutrino in
this setting.

Beyond the specific two-particle calculation, the present analysis illustrates a
more general point: even an extremely weak interaction acquires a clear
quantum-information meaning once its genuinely nonlocal contribution is isolated
from local phase evolution. The self-gravitating neutrino model provides a
controlled analytical setting in which this separation can be carried out
exactly, exposing in closed form the relation between particle mixing,
interaction, and dynamically generated entanglement.

The assumptions adopted here, namely a closed two-particle system, a fixed
interparticle separation, and a Newtonian gravitational potential, are
deliberately restrictive and define the regime in which the analytic treatment
remains transparent. The corresponding time scales should therefore be
interpreted as intrinsic properties of this idealized model rather than as
sensitivity estimates for realistic neutrino experiments. At the same time, the
simplicity of the two-particle solution makes it a natural benchmark for more
general settings. A natural next step is the extension of the analysis to the
$N$-particle gravitational Hamiltonian, in order to determine how bipartite
entanglement is redistributed into genuinely multipartite correlations as the
system size increases. Relaxing the fixed-separation approximation and including
more realistic kinematics would provide a complementary direction for assessing
the robustness of the entangling mechanism beyond the present idealization.
\section*{Acknowledgements}
RS acknowledges support from the project “Implementation of cutting-edge research and its application as part of the Scientific Center of Excellence for Quantum and Complex Systems, and Representations of Lie Algebras“, Grant No. PK.1.1.10.0004, co-financed by the European Union through the European Regional Development Fund - Competitiveness and Cohesion Programme 2021-2027.
A.C.  acknowledges partial support from MIUR and INFN. A.C. also acknowledges COST Action CA1511, Cosmology and Astrophysics Network for
Theoretical Advances and Training Actions (CANTATA).
\bibliographystyle{unsrt_boldvolume}
\bibliography{references}

\end{document}